# Low-temperature synthesis of SmO$_{0.8}$F$_{0.2}$FeAs superconductor with $T_c$ = 56.1 K

Chunlei Wang [a, b], Zhaoshun Gao [a], Lei Wang [a], Yanpeng Qi [a], Dongliang Wang [a], Chao Yao, Zhiyu Zhang [a], Yanwei Ma [a, 1]

[a] Key Laboratory of Applied Superconductivity, Institute of Electrical Engineering, Chinese Academy of Sciences, P. O. Box 2703, Beijing 100190, China
[b] College of physics and electronic engineering, Xinyang Normal University, Xinyang 464000, China

**Abstract**

We report a systematic study on the effect of sintering temperature on the structural and superconducting properties of nominal SmO$_{0.8}$F$_{0.2}$FeAs fabricated by simple one-step solid state reaction method. A detailed correlation between the sintering temperature, structure, onset transition temperature ($T_c$) and critical current density ($J_c$) has been found in all samples of each batch. Most importantly, samples sintered at a low temperature clearly shows high $T_c$, for example the $T_c$ of the samples sintered at 850$^o$C is even above 53 K, and the samples prepared at 1000$^o$C display the highest $T_c$ of 56.1 K reported so far. Furthermore, the samples sintered at 1000$^o$C also show the highest RRR and the lowest $\rho$(57K), indicating the low impurity scattering and enhanced carrier density. However, the maximum $J_c$ of 10510 A/cm$^2$ at 5 K in self field was achieved in the samples sintered at 1100$^o$C. The dependence of $T_c$ on $a$-axis lattice constant indicates that the sintering temperature has strong influence on the effective substitution level of F for O. This result suggests that annealing at a temperature of ~1000$^o$C seems much better for obtaining high quality 1111 phase oxypnictides, compared to commonly used temperatures of around 1200$^o$C.

---

[1] Author to whom any correspondence should be addressed; E-mal: ywma@mail.iee.ac.cn

## 1. Introduction

The discovery of superconductivity in iron-based LaO$_{1-x}$F$_x$FeAs superconductor with an onset transition temperature $T_c$ = 26 K has attracted great interest because of its importance in basic and applied study [1]. Especially, the subsequent experiments indicated that the $T_c$ can exceed 50 K with the replacement of La by other rare earth (RE) elements, such as Sm, Ce, Pr and Nd [2-6]. The rather higher $T_c$ of the oxypnictide superconductors provides a new way for scientists to investigate the mechanisms of superconductivity. For all these REO$_{1-x}$F$_x$FeAs superconductors, recent studies showed that the layered structure with the conducting Fe-As layers should be responsible for the high temperature superconductivity, and the RE-O layers play as a role of charge reservoirs. Most importantly, researchers also found that the newly discovered oxypnictide superconductors usually exhibited exceptionally high upper critical field (H$_{c2}$) [7-12]. For example, over 200 T of upper critical fields has been reported for SmO$_{1-x}$F$_x$FeAs and NdO$_{0.82}$F$_{0.18}$FeAs [8, 10, 13]. The data indicated that these REO$_{1-x}$F$_x$FeAs superconductors should have potential application in high magnetic fields due to the higher values of H$_{c2}$.

In terms of recent reports, the SmO$_{1-x}$F$_x$FeAs superconductors have been successfully prepared by several groups with high-pressure and/or high-temperature sintering methods [3, 5, 14]. Chen et al. [5] firstly synthesized SmO$_{1-x}$F$_x$FeAs by two-step high-temperature sintering process and the highest $T_c$ is only 43 K for the samples sintered at 1160 °C. After that, the $T_c$ has been quickly improved and got up to 55 K by the high-pressure sintering method with pressure as high as 6 GPa at 1250ºC [3]. In addition, our group also developed a simple one-step sintering technique for the preparation of high density SmO$_{1-x}$F$_x$FeAs with the $T_c$ of 52 K [8]. The high sintering temperature can usually be beneficial to the growth of superconducting phase, but it simultaneously causes volatility of some constituents, such as As and F elements, which produces new impurities. Although it has been found that low temperature sintering has some advantages for preparation of MgB$_2$ [15], no similar investigation on SmO$_{1-x}$F$_x$FeAs superconductor has been reported so far. Especially, it is important to study the effect of heating temperatures on the superconducting properties of $SmO_{1-x}F_xFeAs$ to find the optimal preparation conditions. In this work, we will give a detailed research about the influence of sintering temperature on phase formation, $T_c$ and $J_c$ by means of crystal structure, resistivity and magnetic measurements.

## 2. Experimental

The polycrystalline nominal $SmO_{0.8}F_{0.2}FeAs$ bulks were synthesized through a one-step method developed by our group [16]. Stoichiometric amount of $SmF_3$, Fe, and $Fe_2O_3$ powers, Sm filings and As pieces were thoroughly ground by hand with a mortar and pestle. In order to compensate the evaporation of As at high temperatures, extra 10% As was added. The powder was filled into a Ta tube with 8 mm outer diameter and 1 mm wall thickness. After packing, the tube was rotary swaged with final outside diameter of 5.9 mm. It should be pointed out that the grinding and packing processes were carried out in a glove box filled with high purity argon atmosphere. The samples were divided into eight groups according to the final sintering temperatures. The samples were firstly heated at 500 °C for 10 h and then sintered at temperatures ranging from 850 to 1200 °C for 45 hours. In order to reduce oxidation of the samples, high purity argon gas was allowed to flow into the furnace during the sintering processes.

The phase identification was characterized using x-ray diffraction (XRD) analysis with Cu-Kα radiation from 10 to 90°. Lattice constants derived from X-ray diffraction patterns were calculated using an X'Pert Plus program. Resistivity and magnetization measurements were carried out by the standard four-probe method using a Quantum Design PPMS. Critical current densities were deduced from Bean model. Microstructure was determined by scanning electron microscopy (SEM) after peeling away the Ta sheath.

## 3. Results and discussion

Figure 1 shows the x-ray diffraction patterns of $SmO_{0.8}F_{0.2}FeAs$ superconductors which were heat treated at different sintering temperatures, ranging from 850 to 1200°C. As we can see that $SmO_{0.8}F_{0.2}FeAs$ with ZrCuSiAs structure is the main phase for every group of samples and a tiny of amount SmAs and SmOF impurity phases are also detected. The impurity phases of SmOF and SmAs slightly decrease with increasing the sintering temperature and reach the minimum value at about 1000°C for SmOF and 1100°C for SmAs, respectively. The refined lattice parameters are also listed in Table 1. According to the reference [5], the lattice parameters are $a$ = 0.3940 nm and $b$ = 0.8501 nm for F-free sample. The reduction of the lattice parameters in our experiments indicates a successfully chemical substitution of F for O. But the lattice constants are obviously different from samples annealed at different temperatures, which can be attributed to the difference of effective F-substitution level. The $a$-axis lattice constant shows a valley shape dependence on sintering temperature, with a minimum value of 0.39254 nm at 1000°C. On the other hand, for lattice constant $c$, the relationship is similar to that of $a$ when sintering temperature is less than 1000°C, but the change tendency is more complex to some extent when the sintering temperature is greater than 1000°C.

Figure 2 displays the resistivity versus temperature curves (ρ-T) of all samples sintered at different temperatures. It can be seen that all the samples show metallic characteristic before superconducting transition. The samples heated at higher sintering temperatures have relatively lower $T_c$, for example, the $T_c$ is about 50.8 K for the samples sintered at 1200°C. It is surprising to find that the samples sintered at lower sintering temperatures exhibit higher $T_c$, as shown in figure 2, the $T_c$ can get up to 53.5 K even for the samples sintered at 850°C. In addition, it should be pointed out that the samples sintered at 1000°C have the highest onset $T_c$ of 56.1 K, with zero resistivity at 54.1 K, which should be the highest one in Sm-1111 family as far as we know. The onset $T_c$ versus sintering temperature curves are displayed in Figure 3. In order to illustrate the relation between $T_c$ and $a$-axis lattice constant, the dependence of $a$-axis lattice constant versus sintering temperatures is also listed, as shown in figure 3. It can be clearly seen that there is a strong dependence between the $T_c$ and $a$-axis lattice constant. Such a phenomenon can be attributed to the difference of effective F-substitution level among these samples due to different sintering temperatures. The substitution of F for O causes the reduction of lattice constants. Thus, there should be higher F-substitution level in samples with smaller $a$-axis lattice constant. As shown in figure 3, the samples sintered at 1000°C should possess the maximum F$^-$ substitution level, while higher or lower sintering temperature than 1000°C results in the reduction of the F-substitution level in SmO$_{1-x}$F$_x$FeAs. So it can be speculated that the optimal sintering temperature for improving F content in SmO$_{1-x}$F$_x$FeAs superconductor is around 1000°C.

Results obviously show that the samples sintered at 1000°C have the lowest resistivity in the normal state over the entire temperature range compared to other samples in present experiments. The $T_c$ value and residual resistivity ratio, R(300K)/R(57K) (RRR), for the samples were found to be 56.1 K and 6.63, respectively, which are much higher than that obtained from other samples, as shown in table 1. Compared to other samples in the present experiments and previous literature data, low resistivity and high RRR for the samples sintered at 1000°C indicate a lower level of impurity scattering in the samples [3,14]. In addition, based on metallic material Matthiessen's rule [17], the slope of $\alpha = d\rho(T)/dT$ reflects the temperature dependence of lattice scattering. The samples sintered at 1000°C also possess the lowest lattice scattering coefficient ($\alpha$), indicating that there is a relative perfect grain connectivity and a reduction of impurity and void. Furthermore, compared with the samples sintered at 1000°C, we found that the samples sintered at other sintering temperatures have higher impurity scattering and lattice scattering, suggesting that over-high and over-low sintering temperatures produce impurities in the SmO$_{0.8}$F$_{0.2}$FeAs superconductor.

Figure 4 shows the SEM images of SmO$_{0.8}$F$_{0.2}$FeAs, illustrating the change of microstructure with different sintering temperature of 900, 1000, 1100 and 1200°C, respectively. It can be seen that the samples sintered at 900 °C is very porous with large numbers of pores. XRD analysis also reveals that the samples may contain main impurity phases of SmAs and SmOF. The samples become much denser because of the reduction of voids and growth of grains with increasing the sintering temperature from 1000 to 1200°C, as shown in figure 4. In addition, the grain size becomes larger with increasing the heating temperature to 1200°C. In comparison with the samples sintered at 1100 and 1200°C, the samples heated 1000°C have smaller grain sizes and less voids to some extent.

Figure 5 illustrates the magnetic field dependence of the critical current density $J_c$ at 5 K derived from the hysteresis loop width using the extended Bean model of $J_c = 20\Delta M /Va(1-a/3b)$ taking the full sample dimensions. Where $\Delta M$ is the height of magnetization loop, $a$ and $b$ denote the dimensions of the sample perpendicular to the direction of magnetic field ($a < b$). As shown in figure 5, the $J_c$ exhibits strong dependence on the sintering temperature. The $J_c$ increases with increasing the sintering temperature and the maximum $J_c$ is obtained for the samples synthesized at 1100°C. However, further increasing sintering temperature causes the decrease of $J_c$, as shown in figure 5. The samples synthesized with low sintering temperature have lower $J_c$ which can be ascribed to more second phase impurities, voids and loose microstructure. Compared with samples sintered at 1100°C, the slight decrease of $J_c$ for samples sintered at 1150 and 1200°C may be due to the loss of F and As because of burning loss, producing new impurities and more voids and hence resulting in increasing of the percolating current path. Though the samples sintered at 1000°C have the highest $T_c$ (56.1 K), the $J_c$ is relative lower, which probably results from the smaller grain size and the lack of effective pinning center.

The substitution of F for O, which introduces electrons and causes distortions of crystal lattice in Fe-based oxypnictides, plays a key role in superconductivity. It has been reported that optimal doping level in $SmO_{1-x}F_xFeAs$ occurs at $x$=0.2 with the highest $T_c$=54 K [18], however, some people argued that critical transition temperature increases monotonously with increasing the F-doping level, and gets its maximum of 53 K at x=0.4 [19]. Such controversial conclusions provided a clue that the actual substitution level of F for O in $SmO_{1-x}F_xFeAs$ may also be affected by other factors. Our results clearly indicate that there is a strong dependence of the effective F-substitution level on sintering temperature in $SmO_{1-x}F_xFeAs$. The optimal sintering temperature for $SmO_{1-x}F_xFeAs$ takes place at 1000°C with the highest $T_c$=56.1 K, which should be the highest one for $SmO_{1-x}F_xFeAs$ superconductor.

Takahashi et al. have reported that anisotropic shrinkage of lattice constants introduced by external pressure could result in a step increase in onset $T_c$ in F-doped LaOFeAs [20]. Actually, we found that there is an inverse relation between the $T_c$ and *a*-axis lattice constant. The smaller the *a*-axis lattice constant is, the higher the $T_c$ is, as shown in figure 3. The shrinkage of lattice constants induced by either external pressures or by F-doping can enhance charge transfer between the insulating and conducting layers. Furthermore, the samples sintered at 1000°C have the highest $T_c$, RRR and the lowest *a*-axis lattice constant, ρ(57K), suggesting that the samples have the highest actual F-substitution level and carrier density as well as the lowest impurity scattering. Thus, the rise in onset $T_c$ may be a result of the interaction among these factors. It should be also noted that the $J_c$ is almost independent of magnetic field, especially for these samples sintered at higher temperatures. The $J_c$ of samples sintered at 1000°C is only about half of the maximum $J_c$ obtained by the samples sintered at 1100°C, which can be attributed to the smaller grain size and the lack of effective pinning center, as shown in figure 4. Therefore, it is expected that further improvement in transport capability can be achieved through enhancing pinning strengths by lengthening holding time or by introducing defect structures as well as by texturing of $SmO_{1-x}F_xFeAs$.

## 4. Conclusions

The effect of sintering temperature on superconductivity of nominal $SmO_{0.8}F_{0.2}FeAs$ has been investigated. The highest onset transition temperature ($T_c$) can reach 56.1 K for the samples sintered at 1000°C while the maximum critical current density ($J_c$) are observed in the samples sintered at 1100°C. It is surprising to find that there is inverse relationship between the *a*-axis lattice constant and onset transition temperature, which is attributable to the difference of effective F content in the samples sintered at different temperature. Compared with the samples sintered at high temperature (above 1150°C), it is surprising to find the samples sintered at lower sintering temperature (about 1000°C) have the highest F-substitution level, RRR, carrier density and $T_c$. Our results can provide useful information for the preparation of high quality polycrystalline and single crystal $SmO_{1-x}F_xFeAs$ superconductors at significantly lower synthesis temperatures.

**Acknowledgments**

The authors thank Zizhao Gan, Liye Xiao, Haihu Wen and Liangzhen Lin for their help and useful discussion. This work is partially supported by the Beijing Municipal Science and Technology Commission under Grant No. Z07000300700703, National '973' Program (Grant No. 2006CB601004) and National '863' Project (Grant No. 2006AA03Z203).

# Table

Table 1. Lattice constants and RRR for $SmO_{0.8}F_{0.2}FeAs$ samples sintered at different temperatures

| Sintering Temperature (℃) | Lattice parameter constants (nm) | | RRR |
|---|---|---|---|
| | $a$ | $c$ | $\rho(300)/\rho(57)$ |
| 850 | 0.39299 | 0.84986 | 3.028 |
| 900 | 0.39272 | 0.84821 | 4.7098 |
| 950 | 0.39286 | 0.84794 | 4.086 |
| 1000 | 0.39254 | 0.84738 | 6.631 |
| 1050 | 0.39273 | 0.84771 | 4.157 |
| 1100 | 0.39289 | 0.84757 | 3.660 |
| 1150 | 0.39290 | 0.84725 | 3.961 |
| 1200 | 0.39301 | 0.84838 | 2.527 |

# Captions

Fig. 1 XRD pattern of SmO$_{0.8}$F$_{0.2}$FeAs samples synthesized at different sintering temperatures. The impurity phases of SmAs and SmOF are marked by * and #, respectively.

Fig. 2 Temperature dependence of resistivity for SmO$_{0.8}$F$_{0.2}$FeAs samples synthesized at different sintering temperatures from 850 to 1200 $^{\circ}$C.

Fig. 3 The dependence of lattice constant *a* and onset transition temperature on the sintering temperature for SmO$_{0.8}$F$_{0.2}$FeAs samples.

Fig. 4 SEM micrographs of SmO$_{0.8}$F$_{0.2}$FeAs samples synthesized at 900 (a), 1000 (b), 1100 (c), 1200 $^{\circ}$C (d).

Fig.5 Magnetic field dependencies of critical current densities at 5 K for SmO$_{0.8}$F$_{0.2}$FeAs samples sintered at different temperatures.

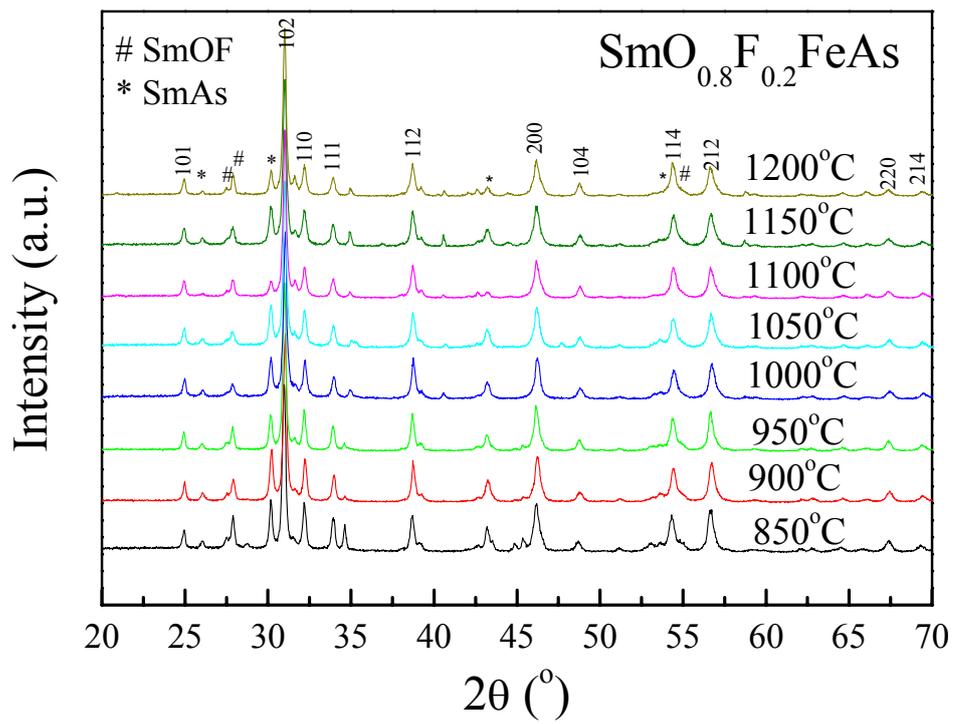

Fig. 1 Wang et al.

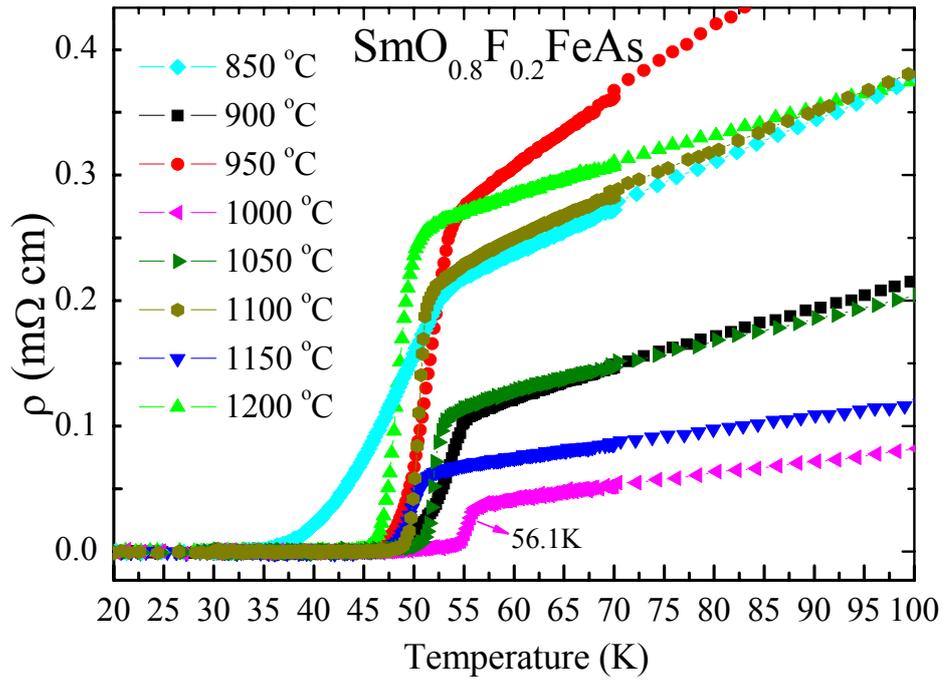

Fig. 2 Wang et al.

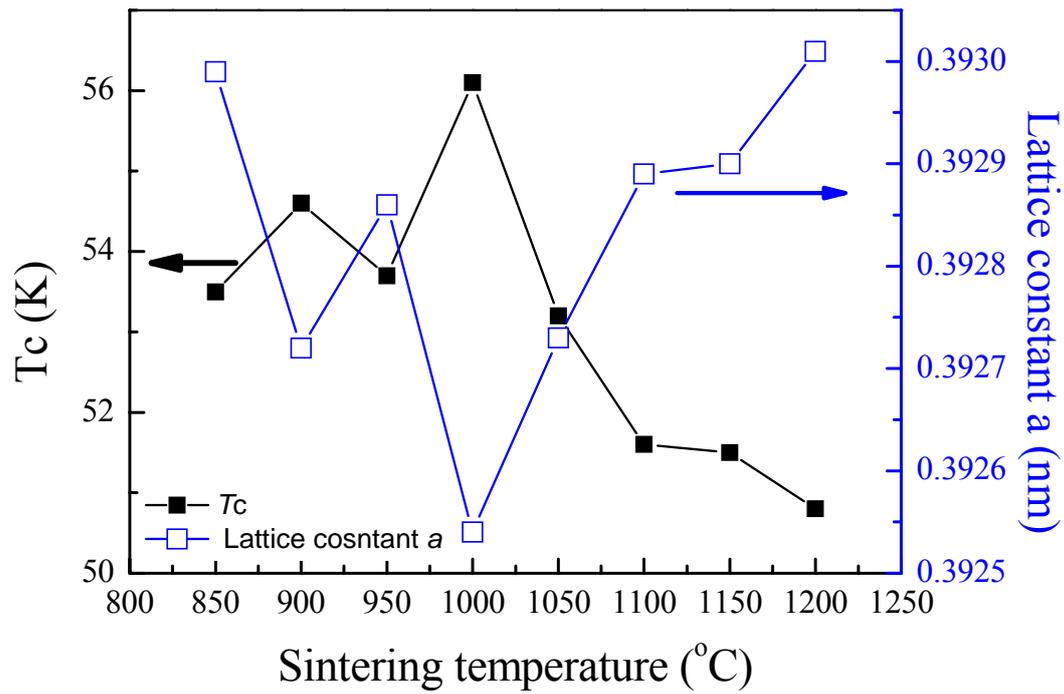

Fig. 3 Wang et al.

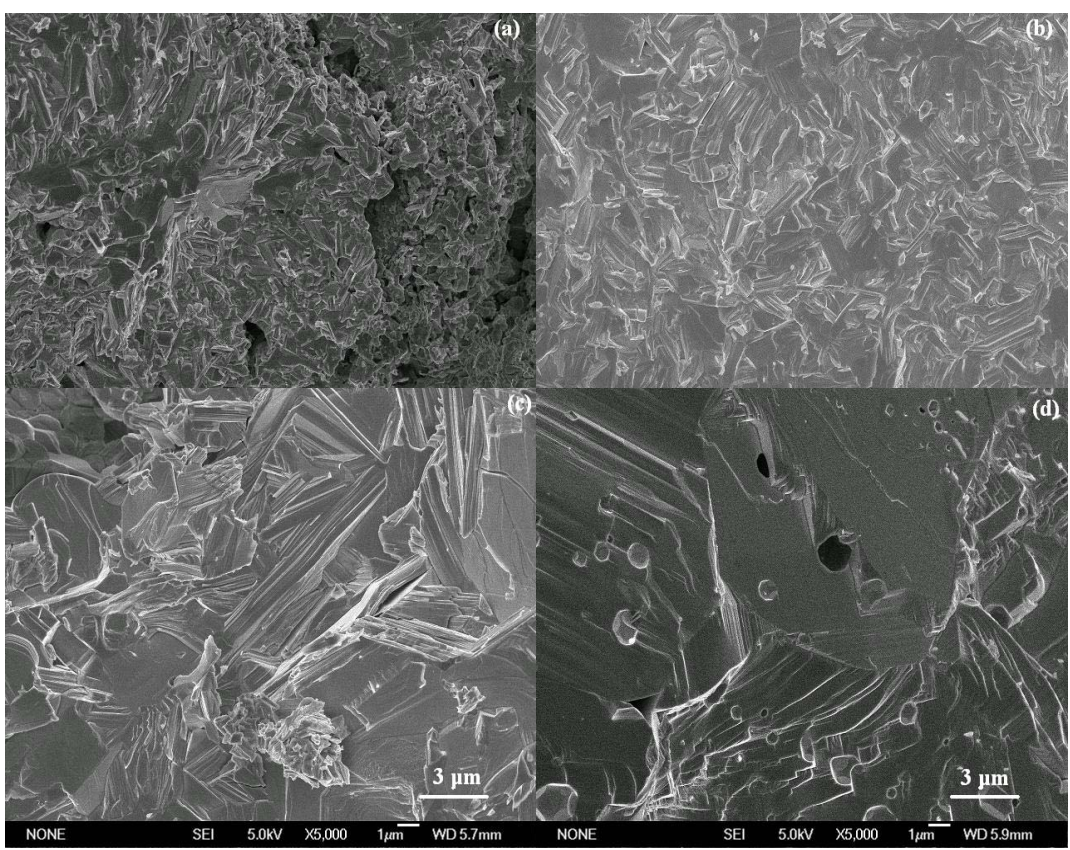

Fig. 4 Wang et al.

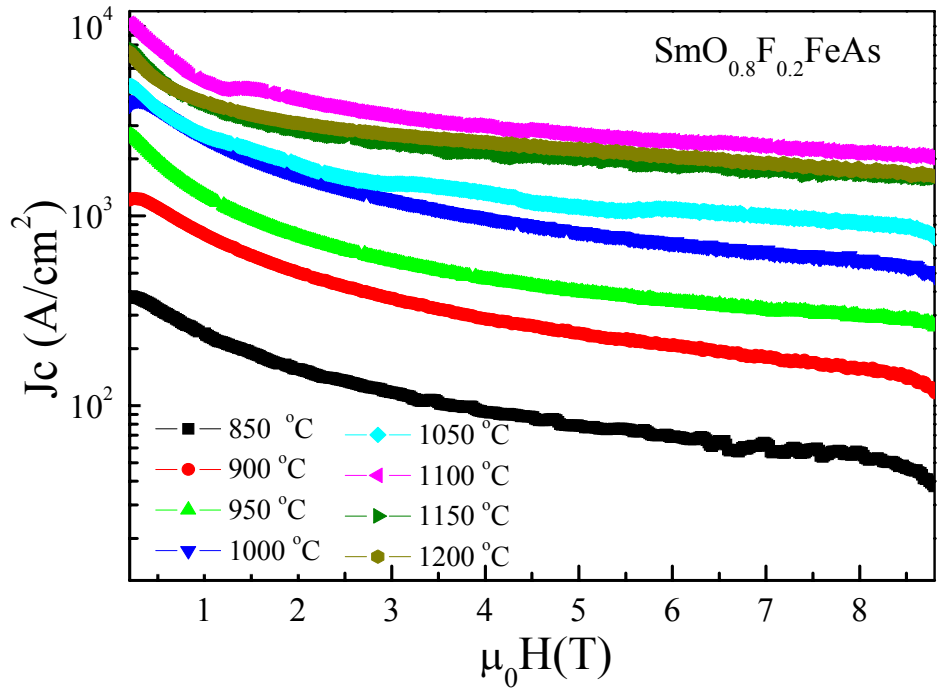

Fig. 5 Wang et al.